\documentstyle[12pt,epsfig]{article}

\newcommand{\re}{\mathop{\rm Re}\nolimits}
\newcommand{\im}{\mathop{\rm Im}\nolimits}

\catcode`@=11
\newcount\@tempcntc
\def\@citex[#1]#2{\if@filesw\immediate\write\@auxout{\string\citation{#2}}\fi
  \@tempcnta\z@\@tempcntb\m@ne\def\@citea{}\@cite{\@for\@citeb:=#2\do
    {\@ifundefined
       {b@\@citeb}{\@citeo\@tempcntb\m@ne\@citea\def\@citea{,}{\bf ?}\@warning
       {Citation `\@citeb' on page \thepage \space undefined}}%
    {\setbox\z@\hbox{\global\@tempcntc0\csname b@\@citeb\endcsname\relax}%
     \ifnum\@tempcntc=\z@ \@citeo\@tempcntb\m@ne
       \@citea\def\@citea{,}\hbox{\csname b@\@citeb\endcsname}%
     \else
      \advance\@tempcntb\@ne
      \ifnum\@tempcntb=\@tempcntc
      \else\advance\@tempcntb\m@ne\@citeo
      \@tempcnta\@tempcntc\@tempcntb\@tempcntc\fi\fi}}\@citeo}{#1}}
\def\@citeo{\ifnum\@tempcnta>\@tempcntb\else\@citea\def\@citea{,}%
  \ifnum\@tempcnta=\@tempcntb\the\@tempcnta\else
   {\advance\@tempcnta\@ne\ifnum\@tempcnta=\@tempcntb \else \def\@citea{--}\fi
    \advance\@tempcnta\m@ne\the\@tempcnta\@citea\the\@tempcntb}\fi\fi}
\catcode`@=12

\begin{document}

\title{\vskip-3cm{\baselineskip14pt
\centerline{\normalsize\hfill NYU--TH/98/05/01}
\centerline{\normalsize\hfill MPI/PhT/98--36}
\centerline{\normalsize\hfill hep--ph/9805390}
\centerline{\normalsize\hfill May 1998}}
\vskip1.5cm
Differences Between the Pole and On-Shell Masses and Widths of the Higgs 
Boson}
\author{{\sc Bernd A. Kniehl}\thanks{Permanent address:
Max-Planck-Institut f\"ur Physik (Werner-Heisenberg-Institut),
F\"ohringer Ring~6, 80805 Munich, Germany.}\ \ {\sc and Alberto Sirlin}\\
{\normalsize Department of Physics, New York University,}\\
{\normalsize 4 Washington Place, New York, NY 10003, USA}}

\date{}

\maketitle

\thispagestyle{empty}

\begin{abstract}
The differences between the on-shell mass and width of the Higgs boson and
their pole counterparts are evaluated in leading order.
For a heavy Higgs boson, they are found to be sensitive functions of the gauge
parameter and become numerically large over a class of gauges that includes
the unitary gauge.
For a light Higgs boson, the differences remain small in all gauges.
The pinch-technique mass and width are found to be close to their pole 
counterparts over a large range of Higgs boson masses.
\medskip

\noindent
PACS numbers: 11.15.Bt, 12.15.Lk, 14.80.Bn 
\end{abstract}

\newpage

\section{Introduction}

The mass and width of an unstable scalar particle are conventionally defined 
by the expressions
\begin{equation}
M^2=M_0^2+\re A(M^2),\qquad
M\Gamma=-\frac{\im A(M^2)}{1-\re A^\prime(M^2)},
\label{eq:os}
\end{equation}
where $M_0$ is the bare mass, $A(s)$ is the self-energy, and the prime 
indicates differentiation with respect to $s$.
Different and, in fact, more fundamental definitions are based on the 
complex-valued position of the propagator's pole:
\begin{equation}
\bar s=M_0^2+A(\bar s).
\end{equation}
Writing $\bar s=m_2^2-im_2\Gamma_2$, in this formulation one may identify the 
mass and width of the unstable particle with $m_2$ and $\Gamma_2$, 
respectively, so that
\begin{equation}
m_2^2=M_0^2+\re A(\bar s),\qquad
m_2\Gamma_2=-\im A(\bar s).
\label{eq:pol}
\end{equation}
Given $m_2$ and $\Gamma_2$, other definitions are possible.
For instance, it has been shown that the alternative expressions
\begin{equation}
m_1=\sqrt{m_2^2+\Gamma_2^2},\qquad
\Gamma_1=\frac{m_1}{m_2}\Gamma_2
\label{eq:lep}
\end{equation}
lead to a Breit-Wigner resonance with an $s$-dependent width and, in the
$Z$-boson case, can be identified with the mass and width measured at LEP
\cite{sir}.
We will refer to Eq.~(\ref{eq:os}) as the on-shell definition of mass and 
width, and to Eq.~(\ref{eq:pol}) or Eq.~(\ref{eq:lep}) as their pole 
counterparts.
Identical formulae hold for spin-1 particles if $A(s)$ is identified with
their transverse self-energy, and analogous expressions can be written down
for spin-1/2 particles.
Most calculations of radiative corrections and widths in the literature employ 
the on-shell formulation of Eq.~(\ref{eq:os}).
On the other hand, in the case of gauge theories, the pole definitions have an 
important advantage:
general arguments imply that the pole position $\bar s$ and, therefore, also
$m_2$, $\Gamma_2$, $m_1$, and $\Gamma_1$ are gauge invariant.
By contrast, it has been shown that the on-shell definitions of $M_W$, $M_Z$, 
and unstable-quark masses become gauge dependent in ${\cal O}(g^4)$ and
${\cal O}(\alpha_sg^2)$ \cite{sir,pas,mpa}.
It has also been pointed out that the on-shell definition of width is 
inadequate if $A(s)$ is not analytic in the neighborhood of $M^2$.
This occurs, for example, when the mass of the decaying particle lies very 
close to a threshold \cite{bha} or, in the resonance region, when the unstable
particle is coupled to massless quanta, such as in the case of the $W$ boson
and unstable quarks \cite{mpa}.

The aim of this letter is to discuss, in leading order, the difference between 
the on-shell mass and width and their pole counterparts for a very important 
case, namely the Higgs boson.
The fact that the width difference may be numerically large for a heavy Higgs 
boson over a large class of gauges is strongly suggested by preliminary 
arguments in Ref.~\cite{rin}.

Expanding Eqs.~(\ref{eq:os}) and (\ref{eq:pol}) about $s=m_2^2$ and combining 
the results, one readily finds
\begin{eqnarray}
\frac{M-m_2}{m_2}&=&-\frac{\Gamma_2}{2m_2}\im A^\prime(m_2^2)+{\cal O}(g^6),
\nonumber\\
\frac{\Gamma-\Gamma_2}{\Gamma_2}&=&
\im A^\prime(m_2^2)\left(\frac{\Gamma_2}{2m_2}+\im A^\prime(m_2^2)\right)
-\frac{m_2\Gamma_2}{2}\im A^{\prime\prime}(m_2^2)+{\cal O}(g^6),
\label{eq:lin}
\end{eqnarray}
where $g^2$ is a generic coupling of ${\cal O}(\Gamma_2/m_2)$.
As the right-hand sides of Eq.~(\ref{eq:lin}) are of ${\cal O}(g^4)$, we may 
evaluate them using the lowest-order expressions for $\Gamma_2$,
$\im A^\prime(m_2^2)$, and $\im A^{\prime\prime}(m_2^2)$.

In the Higgs-boson case, the one-loop bosonic contribution to $\im A(s)$ in
the $R_\xi$ gauge is given by
\begin{eqnarray}
\im A_{\rm bos}(s)&=&\frac{G}{4}s^2\left[-\left(1-\frac{4M_W^2}{s}
+\frac{12M_W^4}{s^2}\right)\left(1-\frac{4M_W^2}{s}\right)^{1/2}
\theta(s-4M_W^2)
\right.\nonumber\\
&&{}+\left.
\left(1-\frac{M_H^4}{s^2}\right)\left(1-\frac{4\xi_WM_W^2}{s}\right)^{1/2}
\theta(s-4\xi_WM_W^2)
+\frac{1}{2}(W\to Z)\right],
\label{eq:bos}
\end{eqnarray}
where $G=G_\mu/(2\pi\sqrt2)$, $\xi_W$ is a gauge parameter, $(W\to Z)$
represents the sum of the preceding terms with the substitutions $M_W\to M_Z$
and $\xi_W\to\xi_Z$, and we have omitted gauge-invariant terms proportional to
$\theta(s-4M_H^2)$.
The one-loop contribution due to a fermion $f$ is
\begin{equation}
\im A_f(s)=-\frac{G}{2}sN_fm_f^2\left(1-\frac{4m_f^2}{s}\right)^{3/2}
\theta(s-4m_f^2),
\label{eq:fer}
\end{equation}
where $N_f=1$ (3) for leptons (quarks).
As expected, Eq.~(\ref{eq:bos}) is gauge invariant if $s=M_H^2$, but it
depends on $\xi_W$ and $\xi_Z$ off-shell.
The $\xi_W$ dependence in Eq.~(\ref{eq:bos}) is due to the fact that a Higgs 
boson of mass $s^{1/2}>2\xi_W^{1/2}M_W$ has non-vanishing phase space to 
``decay" into a pair of ``particles" of mass $\xi_W^{1/2}M_W$.
The first term in Eq.~(\ref{eq:bos}) can be verified by a very simple 
argument \cite{rin}:
only the unphysical scalar excitations have $M_H$-dependent couplings with the
Higgs boson; therefore, if the unphysical particles decouple, which happens
for $\xi_W>s/(4M_W^2)$ and similarly for the $Z$ boson, $\im A(s)$ can be
obtained by substituting $M_H^2\to s$ in the well-known expressions for the
Higgs-boson partial widths multiplied by $M_H$.
Using Eqs.~(\ref{eq:bos}) and (\ref{eq:fer}), we find at the one-loop level:
\begin{eqnarray}
\im A_{\rm bos}^\prime(M_H^2)&=&\frac{G}{2}M_H^2
\left[-\left(1-\frac{5}{4}x_W+\frac{x_W^2}{4}+\frac{3}{16}x_W^3\right)
\left(1-x_W\right)^{-1/2}\theta\left(1-x_W\right)
\right.\nonumber\\
&&{}+\left.
\left(1-\xi_Wx_W\right)^{1/2}\theta\left(1-\xi_Wx_W\right)
+\frac{1}{2}(W\to Z)\right],
\nonumber\\
\im A_{\rm bos}^{\prime\prime}(M_H^2)&=&\frac{G}{2}
\left[-\left(1-\frac{3}{2}x_W+\frac{3}{8}x_W^2-\frac{x_W^3}{4}
+\frac{9}{32}x_W^4\right)
\left(1-x_W\right)^{-3/2}\theta\left(1-x_W\right)
\right.\nonumber\\
&&{}+\left.
\left(1-\xi_Wx_W\right)^{-1/2}\theta\left(1-\xi_Wx_W\right)
+\frac{1}{2}(W\to Z)\right],
\nonumber\\
\im A_f^\prime(M_H^2)&=&-\frac{G}{2}N_fm_f^2\left(1+\frac{x_f}{2}\right)
\left(1-x_f\right)^{1/2}\theta\left(1-x_f\right),
\nonumber\\
\im A_f^{\prime\prime}(M_H^2)&=&-\frac{3}{32}GN_fx_f^3
\left(1-x_f\right)^{-1/2}\theta\left(1-x_f\right),
\label{eq:der}
\end{eqnarray}
where $x_a=4M_a^2/M_H^2$.
Equations~(\ref{eq:bos}), (\ref{eq:fer}), and (\ref{eq:der}) permit us to 
evaluate Eq.~(\ref{eq:lin}).
We also wish to evaluate $(M^{\rm PT}-m_2)/m_2$ and
$(\Gamma^{\rm PT}-\Gamma_2)/\Gamma_2$, where $M^{\rm PT}$ and
$\Gamma^{\rm PT}$ are the pinch-technique (PT) on-shell mass and width 
obtained from Eq.~(\ref{eq:os}) by employing the PT self-energy $a(s)$.
We recall that the PT is a prescription that combines conventional 
self-energies with ``pinch parts" from vertex and box diagrams in such a 
manner that the modified self-energies are independent of $\xi_i$ 
($i=W,Z,\gamma$) and exhibit desirable theoretical properties \cite{cor}.
In the Higgs-boson case, $\im a(s)$ can be extracted from Ref.~\cite{pil},
and we find
\begin{eqnarray}
\im a_{\rm bos}^\prime(M_H^2)&=&\frac{3}{2}GM_W^2
\left(1-x_W-\frac{x_W^2}{4}\right)
\left(1-x_W\right)^{-1/2}\theta\left(1-x_W\right)
+\frac{1}{2}(W\to Z),
\nonumber\\
\im a_{\rm bos}^{\prime\prime}(M_H^2)&=&\frac{G}{4}x_W
\left(1+\frac{x_W}{4}-\frac{x_W^2}{2}-\frac{9}{16}x_W^3\right)
\left(1-x_W\right)^{-3/2}\theta\left(1-x_W\right)
+\frac{1}{2}(W\to Z).
\end{eqnarray}

\begin{figure}[ht]
\begin{center}
\epsfig{figure=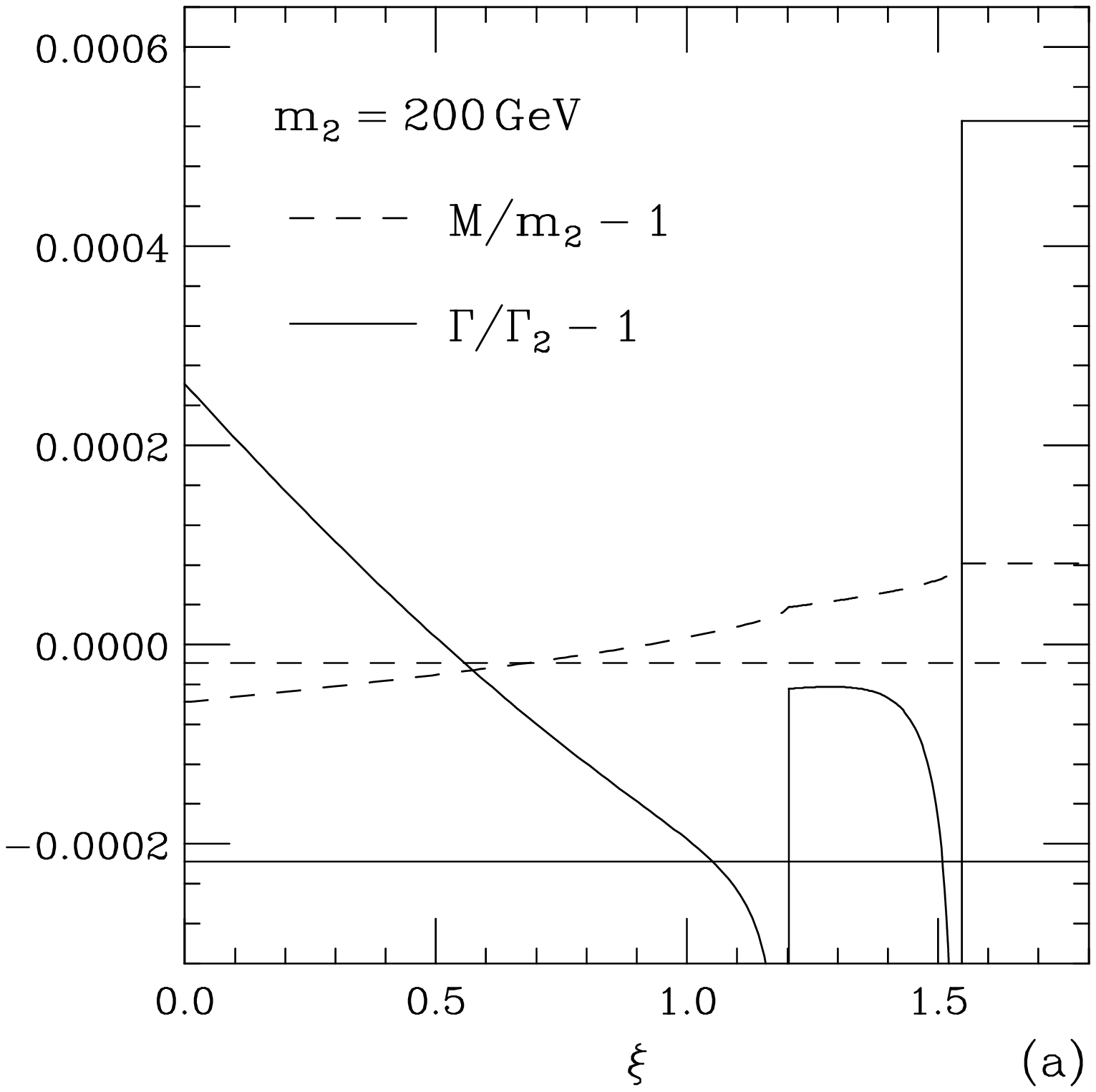,width=\textwidth}
\caption{Relative deviations of $M$ and $\Gamma$ from $m_2$ and $\Gamma_2$, 
respectively, as functions of $\xi=\xi_W=\xi_Z$ in the $R_\xi$ gauge, assuming
(a) $m_2=200$~GeV, (b) 400~GeV, and (c) 800~GeV.
The horizontal lines across the figures indicate the corresponding deviations
in the PT framework.}
\label{fig:one}
\end{center}
\end{figure}

Identifying $M_H$ with $m_2$ and, for simplicity, setting $\xi=\xi_W=\xi_Z$, 
our results for $(M-m_2)/m_2$ and $(\Gamma-\Gamma_2)/\Gamma_2$ are illustrated
in Figs.~\ref{fig:one}(a)--(c) as functions of $\xi$, for three values of 
$m_2$.
We have employed $M_W=80.375$~GeV, $M_Z=91.1867$~GeV, and $m_t=175.6$~GeV, and
have neglected contributions from fermions other than the top quark.
The two deep abysses in the figures are associated with the unphysical 
thresholds $\xi=m_2^2/(4M_Z^2),m_2^2/(4M_W^2)$, where the expansions in 
Eq.~(\ref{eq:lin}) obviously fail.
For small Higgs mass ($m_2=200$~GeV), we see from Fig.~\ref{fig:one}(a) that,
aside from the neighborhoods of the abysses, $M$ and $\Gamma$ remain 
numerically very close to $m_2$ and $\Gamma_2$.
In the intermediate case ($m_2=400$~GeV), the relative differences reach
0.6\% in the mass and 3.3\% in the width.
However, for a heavy Higgs boson ($m_2=800$~GeV), the differences become very
large, reaching 11\% in the mass and 44\% in the width.
The largest differences occur for $\xi>m_2^2/(4M_W^2)$, i.e., when the
unphysical excitations decouple, a range that includes the unitary gauge.
We recall that the latter retains only the physical degrees of freedom and, in 
this sense, it may be regarded as the most physical of all gauges.
The large effects can be easily understood from Eq.~(\ref{eq:bos}).
If $\xi>s/(4M_W^2)$, the second term in Eq.~(\ref{eq:bos}) does not
contribute, so that $\im A_{\rm bos}(s)\propto s^2$.
For a heavy Higgs boson, this implies large values of $\im A^\prime(m_2^2)$
and $\im A^{\prime\prime}(m_2^2)$.
For $\xi<s/(4M_Z^2)$, the gauge-dependent terms contribute and cancel the 
leading $s^2$ dependence of $\im A_{\rm bos}(s)$, so that the magnitudes of
$\im A^\prime(m_2^2)$ and $\im A^{\prime\prime}(m_2^2)$ drop sharply and the 
differences become much smaller.
Of course, the 44\% effect in the width for $\xi>m_2^2/(4M_W^2)$ may cast 
doubts on the convergence of the expansions in Eq.~(\ref{eq:lin}).
We interpret this finding as an indication of large corrections rather than a 
precise evaluation of $(\Gamma-\Gamma_2)/\Gamma_2$.

Our results go beyond those reported in the literature \cite{wil}.
The reason is easy to understand:
in Ref.~\cite{wil}, the limits $M_W\to0$ and $g\to0$ are simultaneously 
considered keeping the Higgs self-coupling $\lambda\propto g^2M_H^2/M_W^2$ 
fixed.
If the gauge parameter $\xi$ is also kept fixed, the gauge dependence of 
Eq.~(\ref{eq:bos}) is lost, and one obtains an $s$-independent result for
$\im A_{\rm bos}(s)$, which does not contribute to the right-hand sides of
Eq.~(\ref{eq:lin}).
Thus, the above approximation, although interesting and useful, does not
exhibit the gauge dependence and the large effects discussed here.

From the horizontal lines across Figs.~\ref{fig:one}(a)--(c), we see that the 
PT mass and width remain very close to $m_2$ and $\Gamma_2$ for all values of
$m_2$, the maximum departures being 0.7\% for $M^{\rm PT}$ and $-0.7\%$ for
$\Gamma^{\rm PT}$ at $m_2=800$~GeV.
The differences vary somewhat if $M$ and $\Gamma$ are compared with $m_1$ and
$\Gamma_1$.
Through ${\cal O}(g^6)$, $(M-m_1)/m_1$ and $(\Gamma-\Gamma_1)/\Gamma_1$ are 
obtained from $(M-m_2)/m_2$ and $(\Gamma-\Gamma_2)/\Gamma_2$ by subtracting
the gauge-invariant term $\Gamma_2^2/(2m_2^2)$.
For $m_2=800$~GeV, $(M-m_1)/m_1$ and $(\Gamma-\Gamma_1)/\Gamma_1$ amount to
5.6\% and 38\% in the unitary gauge (rather than 11\% and 44\%) and to
$-4.8\%$ and $-6.6\%$ in the 't~Hooft-Feynman gauge (rather than 0.9\% and 
$-0.8\%$).
For the same value of $m_2$, the differences $(M^{\rm PT}-m_1)/m_1$ and
$(\Gamma^{\rm PT}-\Gamma_1)/\Gamma_1$ are $-5.1\%$ and $-6.5\%$ (rather than
0.7\% and $-0.7\%$).

In summary, we have shown that, in leading order, the differences between the 
on-shell mass and width of a heavy Higgs boson and their pole counterparts are
sensitive functions of the gauge parameter, and reach large numerical values
in a class of gauges that includes the unitary gauge.
For other frequently employed gauges, such as $\xi=1$ ('t~Hooft-Feynman gauge)
and $\xi=0$ (Landau gauge), the differences are very small with respect to
$m_2$ and $\Gamma_2$, but are not negligible relative to $m_1$ and $\Gamma_1$.
For intermediate (light) Higgs bosons, the differences are reasonably (very) 
small for all values of $\xi$, except in the abysses described above.
The PT on-shell mass and width remain close to $m_2$ and $\Gamma_2$ in the
range 200~GeV${}\le m_2\le800$~GeV.
These results give further support to the proposition that a consistent 
definition of two of the most important concepts in particle physics, namely 
those of mass and width of an unstable particle, must ultimately be based on 
the pole position rather than the on-shell approach 
\cite{sir,pas,mpa,bha,rin,mco}.
For many purposes, the well-known and convenient machinery of the latter can 
be employed, but physicists should become aware of its limitations and 
potential pitfalls.

\vspace{1cm}
\noindent
{\bf Acknowledgements}
\smallskip

\noindent
B.A.K. thanks the NYU Physics Department for the hospitality extended to him 
during a visit when this manuscript was prepared.
This research was supported in part by NSF Grant No.\ PHY--9722083.

\newpage

\newpage
\begin{figure}[ht]
\begin{center}
\epsfig{figure=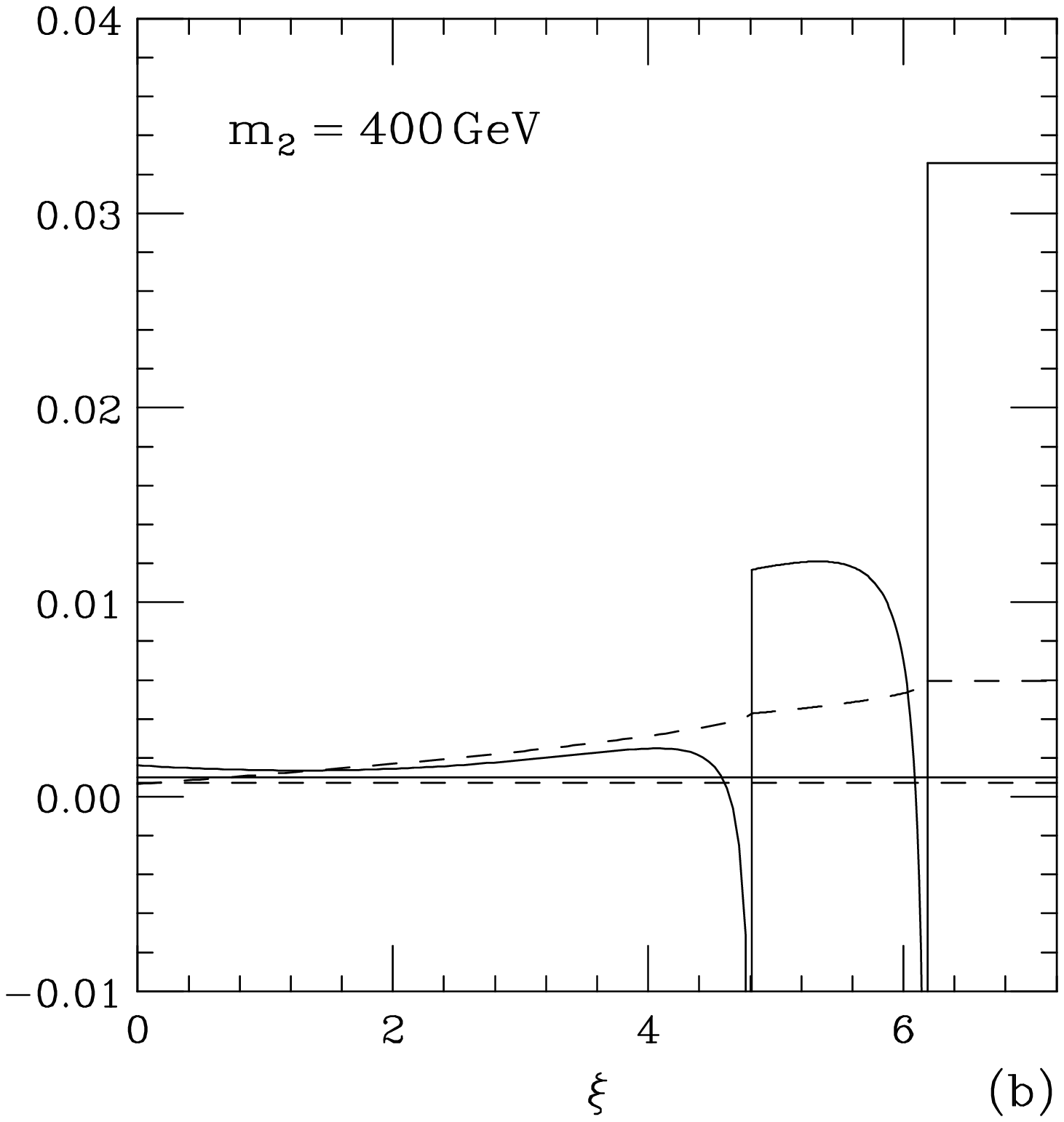,width=\textwidth}
\end{center}
\end{figure}

\newpage
\begin{figure}[ht]
\begin{center}
\epsfig{figure=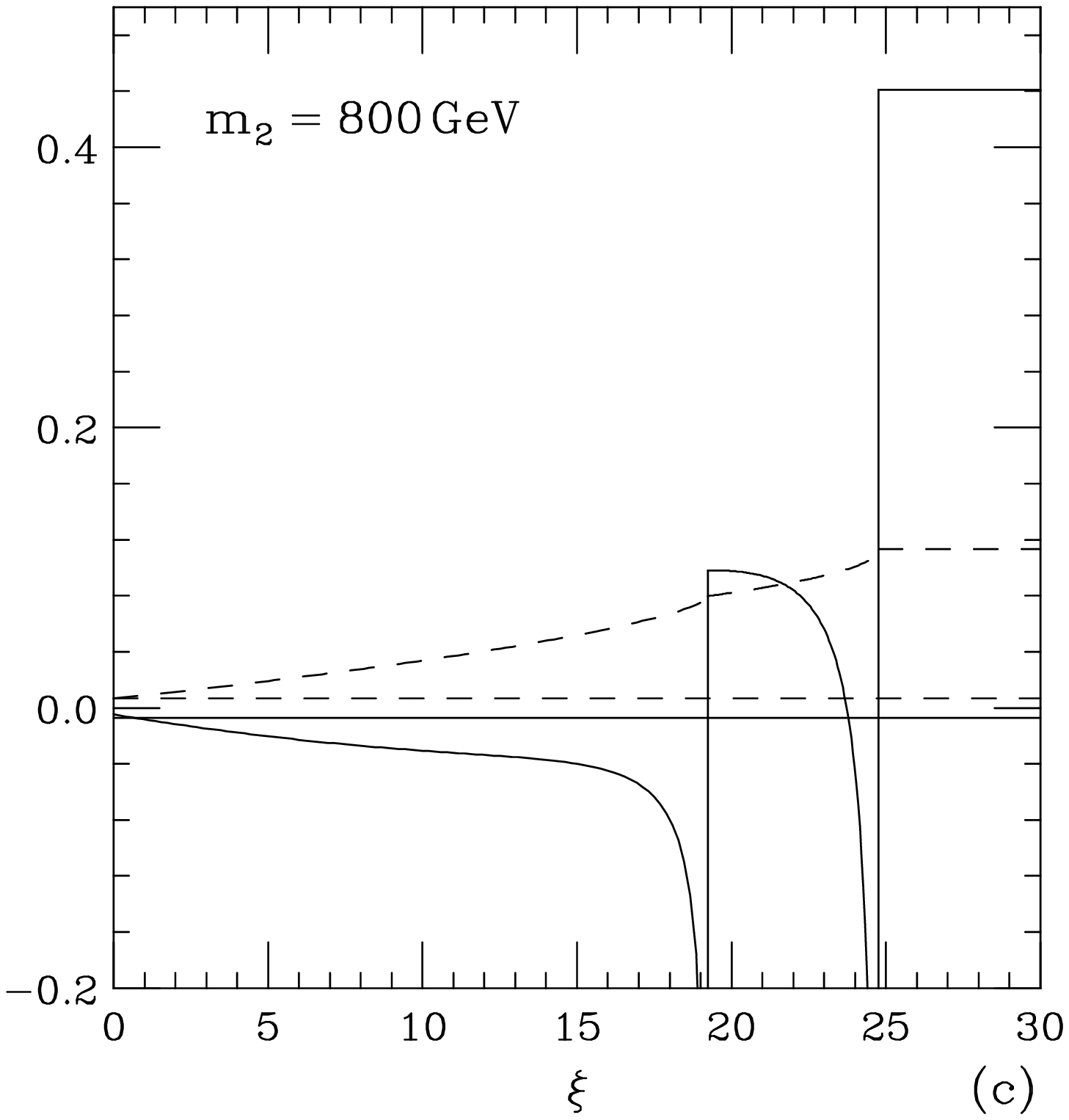,width=\textwidth}
\end{center}
\end{figure}

\end{document}